\documentclass[useAMS,usenatbib]{mn2e}
\usepackage{graphicx,amssymb,color}
\usepackage[normalem]{ulem}

\title[Constraints on co-orbital bodies]
{Dynamical mass and multiplicity constraints on co-orbital bodies around stars}

\author[Veras, Marsh \& G\"{a}nsicke]{
Dimitri Veras$^{1}$\thanks{E-mail: d.veras@warwick.ac.uk},
Thomas R. Marsh$^{1}$,
Boris T. G\"{a}nsicke$^{1}$
\\
$^{1}$Department of Physics, University of Warwick, Coventry CV4 7AL, UK
}

\date{Accepted 2016 May 30. Received 2016 May 3; in original form 2016 February 22}
\pubyear{2016}

\begin{document}
\label{firstpage}
\pagerange{\pageref{firstpage}--\pageref{lastpage}}
\maketitle

\begin{abstract}
Objects transiting near or within the disruption radius of both main sequence 
(e.g. KOI~1843) and white dwarf (WD 1145+017) stars are now known. Upon fragmentation or disintegration, 
these planets or asteroids may produce co-orbital configurations of nearly equal-mass objects.
However, as evidenced by the co-orbital objects detected by transit photometry in the WD 1145+017 system, 
these bodies are largely unconstrained in size, 
mass, and total number (multiplicity). Motivated by potential future similar discoveries, we perform $N$-body 
simulations to demonstrate if and how debris masses and multiplicity may be bounded due to second-to-minute 
deviations and the resulting accumulated phase shifts in the osculating orbital period amongst multiple 
co-orbital equal point masses.  We establish robust lower and upper mass bounds as a function of orbital period 
deviation, but find the constraints on multiplicity to be weak. We also quantify the fuzzy instability 
boundary, and show that mutual collisions occur in less than  5\%, 10\% and 20\% of our simulations for 
masses of $10^{21}$, $10^{22}$ and $10^{23}$ kg.  Our results may provide useful initial rough constraints 
on other stellar systems with multiple co-orbital bodies.
\end{abstract}


\section{Introduction}

One recent exciting development in exoplanetary science has been the discovery
of actively dying planets and asteroids.  The main sequence stars KOI~2700, KIC~12557548B, 
and K2-22 all contain planet candidates which are thought to be disintegrating 
due to dusty flows or tails \citep{rapetal2012,rapetal2014,croetal2014,bocetal2015,sanetal2015}.
These systems harbour objects with orbital periods of, respectively, about 22, 16 and 9 hours.
The KOI~1843 system contains a planet candidate which, although not yet observed
to be disintegrating, is close enough to its star's disruption, or Roche, radius, for a lower 
bound on its density to be set at 7 g cm$^{-3}$ \citep{rapetal2013}.
This object's orbital period is only 4.245 hours.

These four systems provide just a taste of the widespread disruption which is 
assumed to occur around stars which have left the main sequence.  
In particular, between one quarter and one half of
all known Milky Way white dwarfs possess atmospheres which 
are ``polluted'' from the metal-rich remnants of shorn-up planetary systems 
\citep{zucetal2003,zucetal2010,koeetal2014}.
The elemental profile of the pollution provides unique insight into planet formation and bulk
chemical composition \citep{zucetal2007,kleetal2011,ganetal2012,xuetal2014,juryou2014,wiletal2015,wiletal2016}, 
and accretion from dusty and sometimes gaseous debris discs likely gives rise to the 
pollution \citep{zucbec1987,gaeetal2006,faretal2009,beretal2014,wiletal2014,baretal2016,farihi2016,manetal2016}.  
The dynamical origin of these pollutants and their pathways through 
all stages of stellar evolution remains uncertain and represents a growing field of
exploration \citep{veras2016}.

Although the tidal disruption of asteroids which veer into the Roche radius of the
white dwarf has long been theorised to represent the dominant polluting mechanism 
\citep{graetal1990,jura2003,bonetal2011,debetal2012,beasok2013,frehan2014}, visual confirmation of this process was not
supplied until the discovery of transiting bodies orbiting WD\,1145+017 \citep{vanetal2015}.
Plentiful and ongoing follow-up observations of this system 
\citep{croetal2015,gaeetal2016,rapetal2016,aloetal2016,xuetal2016,zhoetal2016} 
showcase complex dynamical signatures amongst at least six objects with orbital periods directly
measured from photometric transit curves, sometimes with individual uncertainties as small as a 
few seconds.  For main sequence stars, although co-orbital solid bodies have not yet been observed, 
an episode of catastrophic fragmentation might produce such bodies, leading to a similar architecture.

Such configurations raise fundamental questions about orbital dynamics, and about how mutual
gravitational interactions may be linked to observations.  Exquisite orbital period
measurements belie the otherwise starkly unconstrained state of such systems: 
unknowns include the size, mass and multiplicity of the co-orbiting objects 
(henceforth referred to as just ``bodies''), as well as the star's 
mass.  For a system like WD 1145+017, the size and mass of the bodies are unconstrained to 
within many orders of magnitude,
whereas, because the star is a white dwarf, its mass is most likely to lie in the range 
$0.5M_{\odot}-0.7M_{\odot}$ (e.g. Fig. 1 of \citealt*{koeetal2014}, and \citealt*{treetal2016}).  
Regardless of these
uncertainties, transit signatures strongly suggest that dust
is emanating from the bodies, and Roche radius computations 
\citep[e.g. based on equations and discussion from][]{murder1999,corsha2008,veretal2014a,beasok2015} affirm
that the bodies are highly likely to be currently disintegrating.

Despite the unknowns, mutual interactions among the co-orbiting bodies can provide some theoretical constraints.  
These interactions will cause deviations in orbital
period and ensuing accumulated phase shifts, both of 
which may be measurable. In this paper, we constrain the masses and 
multiplicities of 
co-orbital bodies in compact configurations by computing orbital period deviations with $N$-body numerical 
simulations. Although we use WD\,1145+017 as inspiration, we do not attempt to specifically model this system
because tidal disruption \citep{debetal2012,veretal2014a},
interaction with the extant disc \citep{rafikov2011a,metetal2012,rafikov2011b,rafgar2012}
and rotational and orbital evolution due to white dwarf radiation 
\citep{veretal2014b,veretal2015a,veretal2015b,veretal2015c,stoetal2015} all likely play a role. 

Instead, we perform simulations with an eye for future observations of similar systems
around any type of star, in order to provide investigators with a basic notion of 
how orbital period deviations correspond
to different architectures without complexities beyond point-mass gravitational dynamics.  
In Section 2, we briefly review co-orbital point mass dynamics with one central massive
body. Section 3 presents our simulation setup and Section 4 displays our results.
We conclude in Section 5.

\section{Co-orbital dynamics}

For nearly 250 years, researchers have attempted to understand how multiple objects 
may share the same orbit around a more massive primary \citep[e.g.][]{lagrange1772}.
A significant initial focus was the three-body problem, through which analytical
stability formulae \citep{gascheau1843,routh1875} proved reliable after the discovery
of Trojan asteroids \citep{wolf1906} and the only known pair of co-orbital
satellites, Janus \citep{dollfus1967} and Epimetheus \citep{foular1978}.
The more complex $N$-body co-orbital problem with $N > 3$ features a larger
phase space, but has been studied primarily since the pioneering work of \cite{maxwell1890}.
He showed that in the limit of large $N$ and for low-enough masses, stable rings may be achieved.
For finite $N$, however, even if the bodies are symmetrically spaced, 
predicting the stability of the system becomes nontrivial \citep{pendse1935,salyod1988}.

Effectively, the case $N > 3$ requires numerical simulations unless the system
in question can be linked to central configurations \citep{moeckel1994,rensic2004},
a multi-body hierarchical restricted problem -- which can be expressed entirely in terms
of orbital element equations of motion -- \citep{veras2014a}, or specialised symmetric 
cases \citep[e.g.][]{benetal2015}.  Few- or many-body co-orbital
dynamics may also be informed by the periodic orbits of the $N=3$ case
\citep{hadetal2009,hadvoy2011,antetal2014}.

The discovery of extrasolar planets \citep{wolfra1992,wolszczan1994,mayque1995}
prompted a resurgence of interest in the co-orbital problem.  Despite the absence 
of discoveries of Trojan planets around main sequence stars (but see \citealt*{gozkon2006}),
the problem has received renewed attention in terms of planet formation and evolution 
\citep{nauenberg2002,koretal2004,schetal2005,beaetal2007,crenel2009,izoetal2010,smilis2010,robpou2013,pieray2014},
Doppler radial velocity detections \citep{laucha2002,giuetal2012,dobrovolskis2013,leletal2015}, and 
detections by transit photometry
\citep{forgau2006,forhol2007,janson2013,voknes2014,plaetal2015} and binary eclipses \citep{schetal2015}.

This work differs from all of the above investigations due to the heretofore unforeseen
character of the transit observations of polluted white dwarfs with disintegrating bodies
(at least, as evidenced by WD\,1145+017).  Because we have no reason to believe that the
bodies should be symmetrically spaced (see e.g. Fig. 3 of \citealt*{gaeetal2016} and 
Fig. 6 of \citealt*{rapetal2016}) -- despite this configuration 
being attractive analytically -- 
we use $N$-body integrations to explore non-symmetrically-spaced bodies 
(along the same orbit).  Our simulations primarily yield orbital period 
variations which can be compared to observations.

\begin{figure*}
\centerline{
\includegraphics[width=8cm,height=6.8cm]{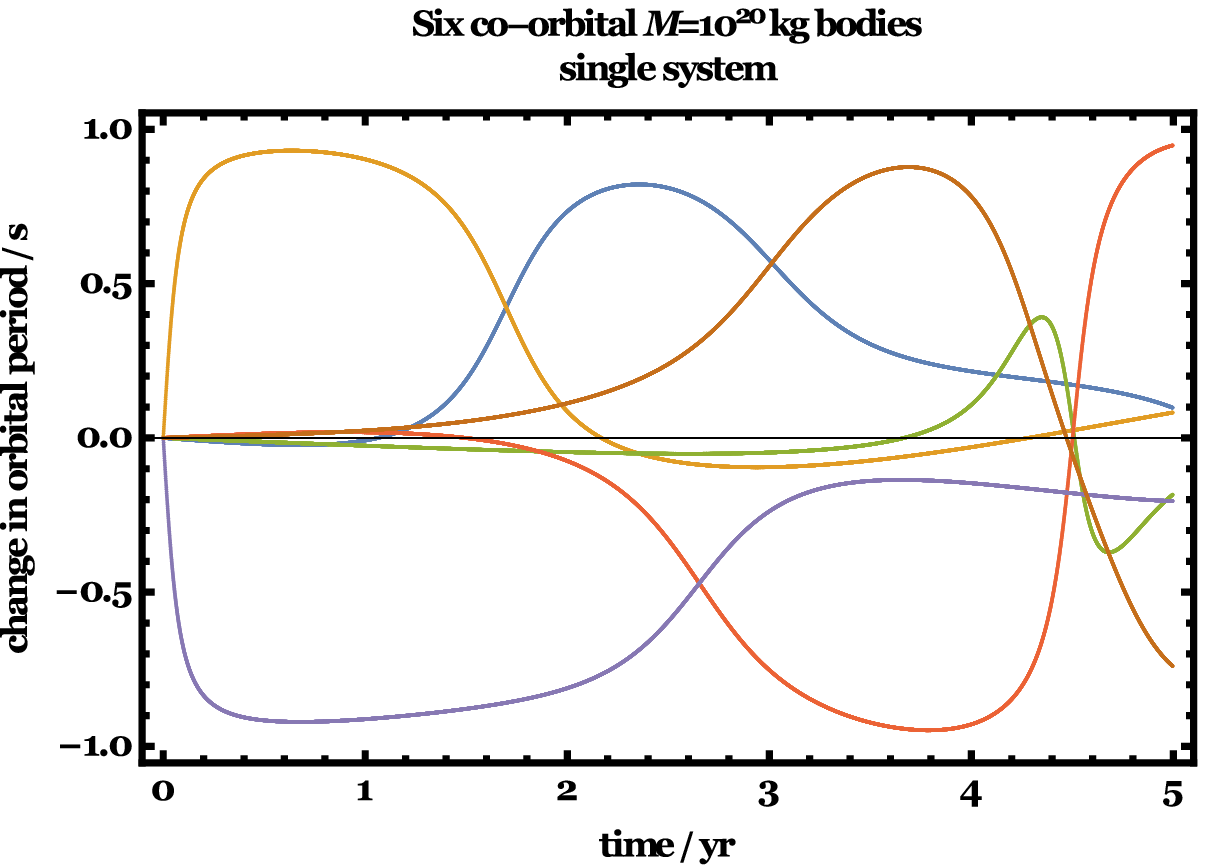}
\hspace{0.20em}
\includegraphics[width=8cm,height=6.8cm]{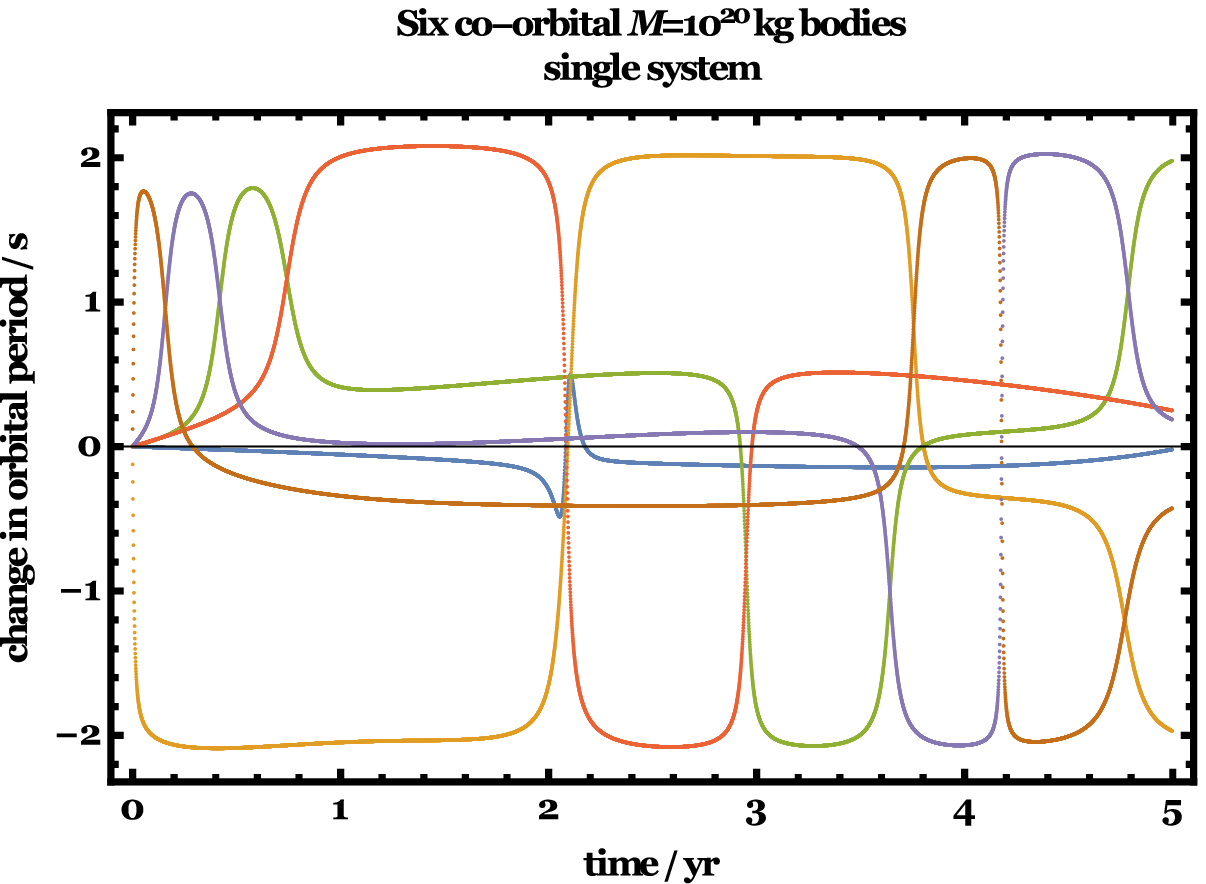}
}
\vspace{-0.00em}
\centerline{
\includegraphics[width=8cm,height=6.8cm]{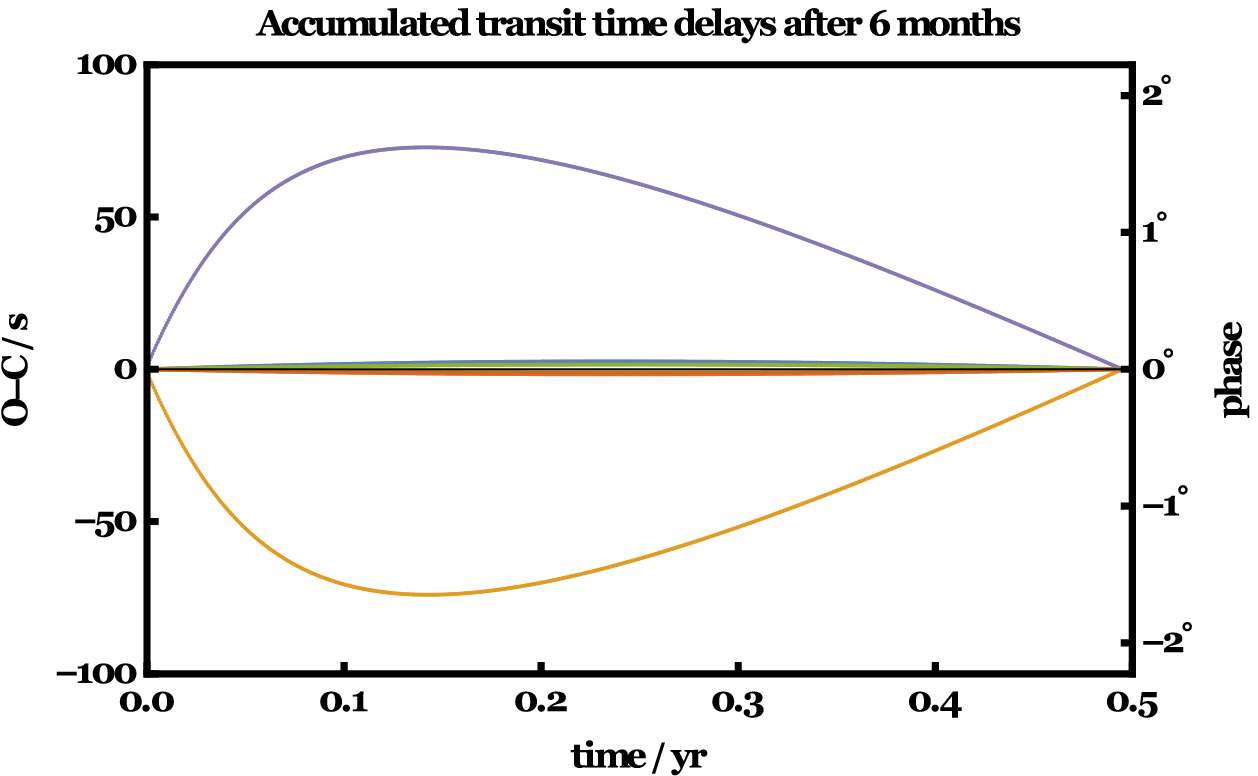}
\hspace{0.20em}
\includegraphics[width=8cm,height=6.8cm]{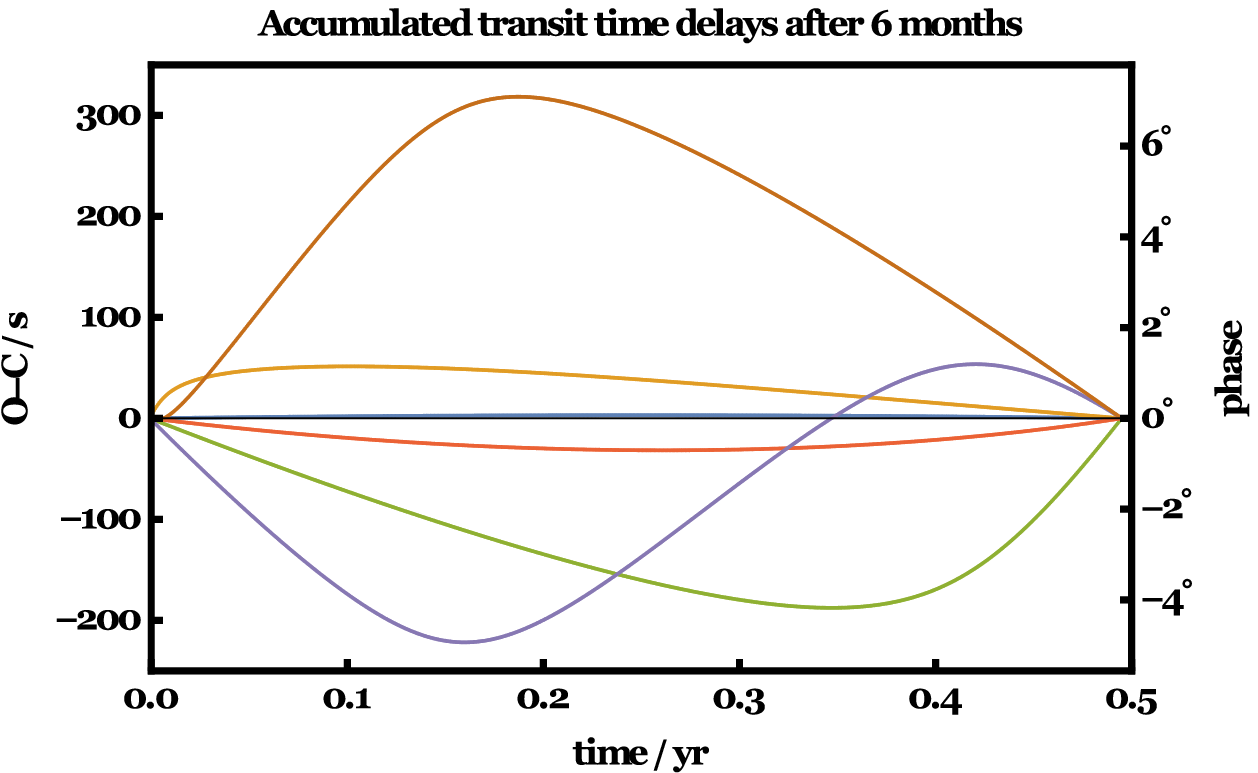}
}
\vspace{-0.00em}
\centerline{
\includegraphics[width=8cm,height=6.8cm]{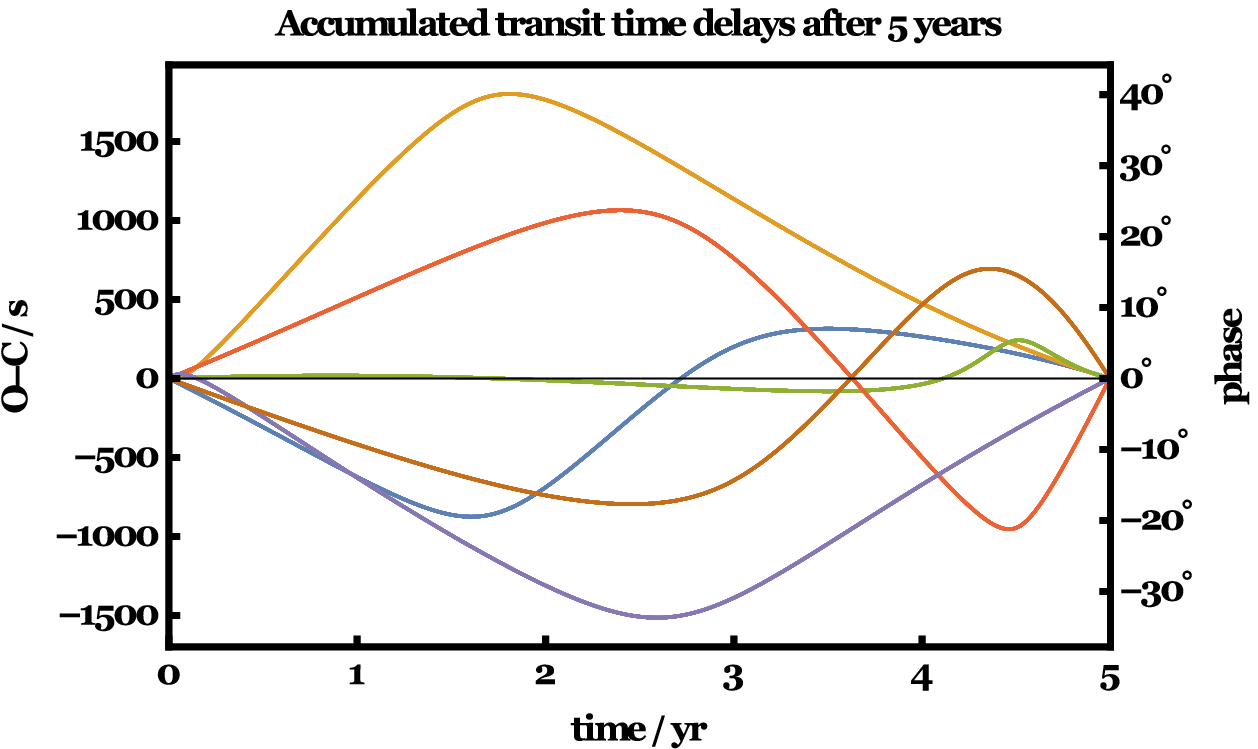}
\hspace{0.20em}
\includegraphics[width=8cm,height=6.8cm]{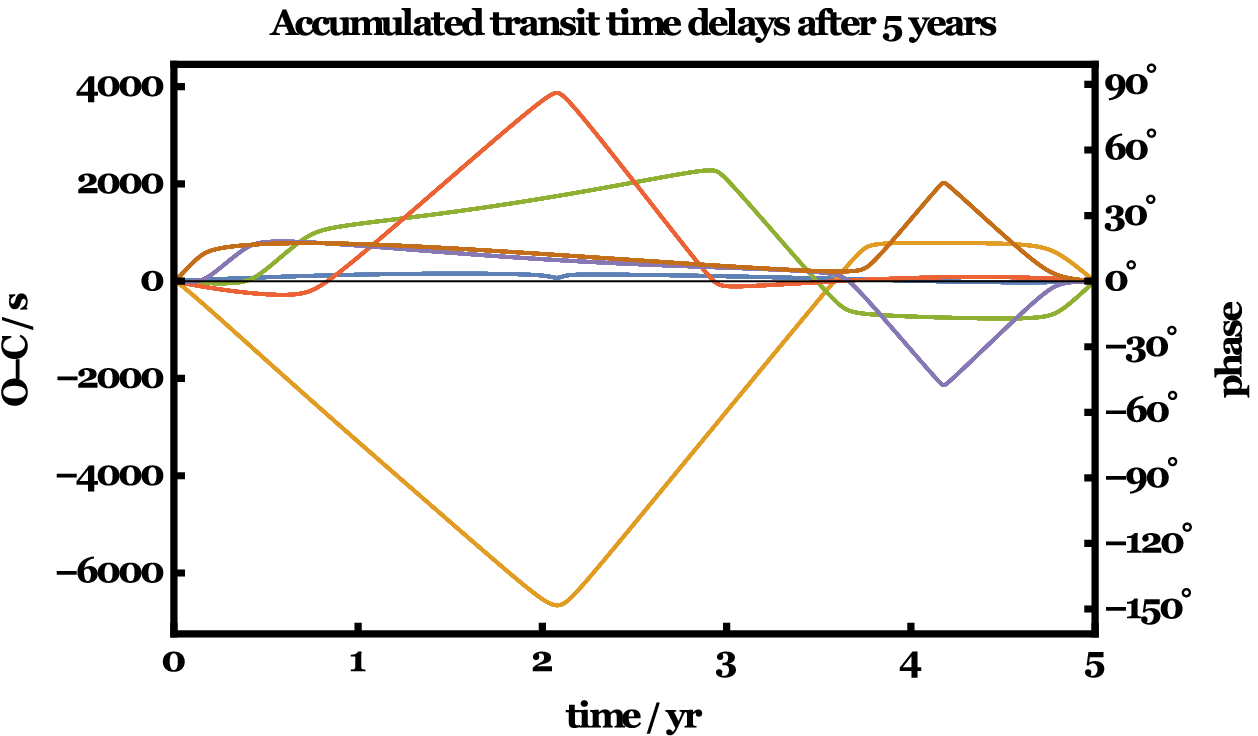}
}
\caption{
The variation in orbital period ({\it upper panels}) and O-C 
(observed - calculated)
deviations of
transit dip times from a linear ephemeris ({\it middle
and lower panels}) for two systems of six $10^{20}$ kg co-orbital bodies
whose orbits share an initial period of 4.49300 hours.  The initial
mean anomalies for all simulations in this paper were selected
from a uniform random distribution; shown here are two cases with clustered
sets of initial mean anomalies of about {\it left panels}:
$176.3^{\circ}$, $219.4^{\circ}$, $238.5^{\circ}$, 
$253.1^{\circ}$, $262.3^{\circ}$ and $292.7^{\circ}$, and {\it right panels}:
$121.0^{\circ}$, $251.9^{\circ}$, $273.9^{\circ}$, $293.1^{\circ}$, $304.7^{\circ}$
and $305.6^{\circ}$. Despite the similarity in these number sets, the
left and right panels exhibit distinctive behaviour and amplitudes.
Both systems would be detectable with current technology.
}
\label{linepl}
\end{figure*}

\begin{figure}
\centerline{
\includegraphics[width=7.5cm,height=6.8cm]{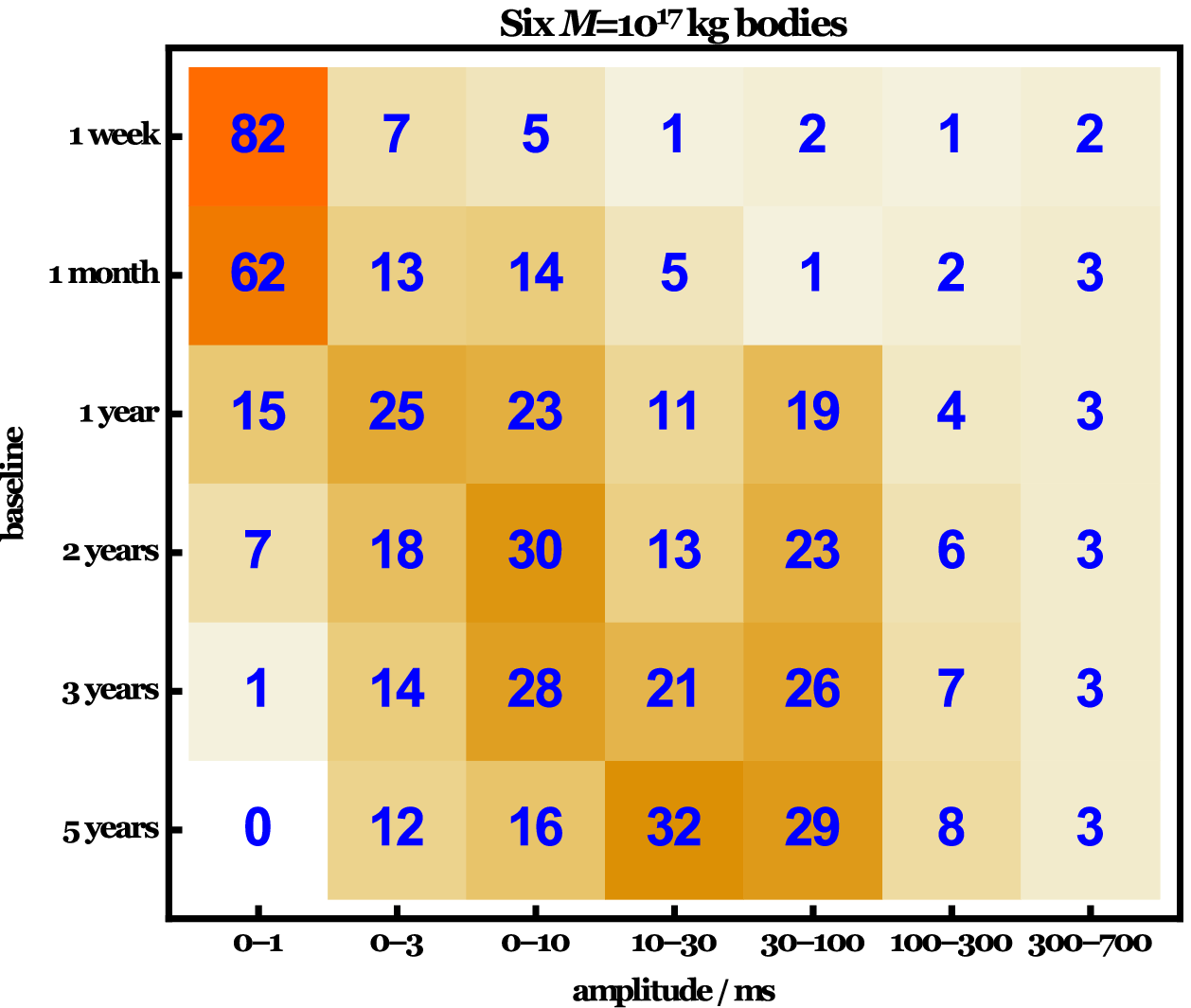}
}
\vspace{0.40em}
\centerline{
\includegraphics[width=7.5cm,height=6.8cm]{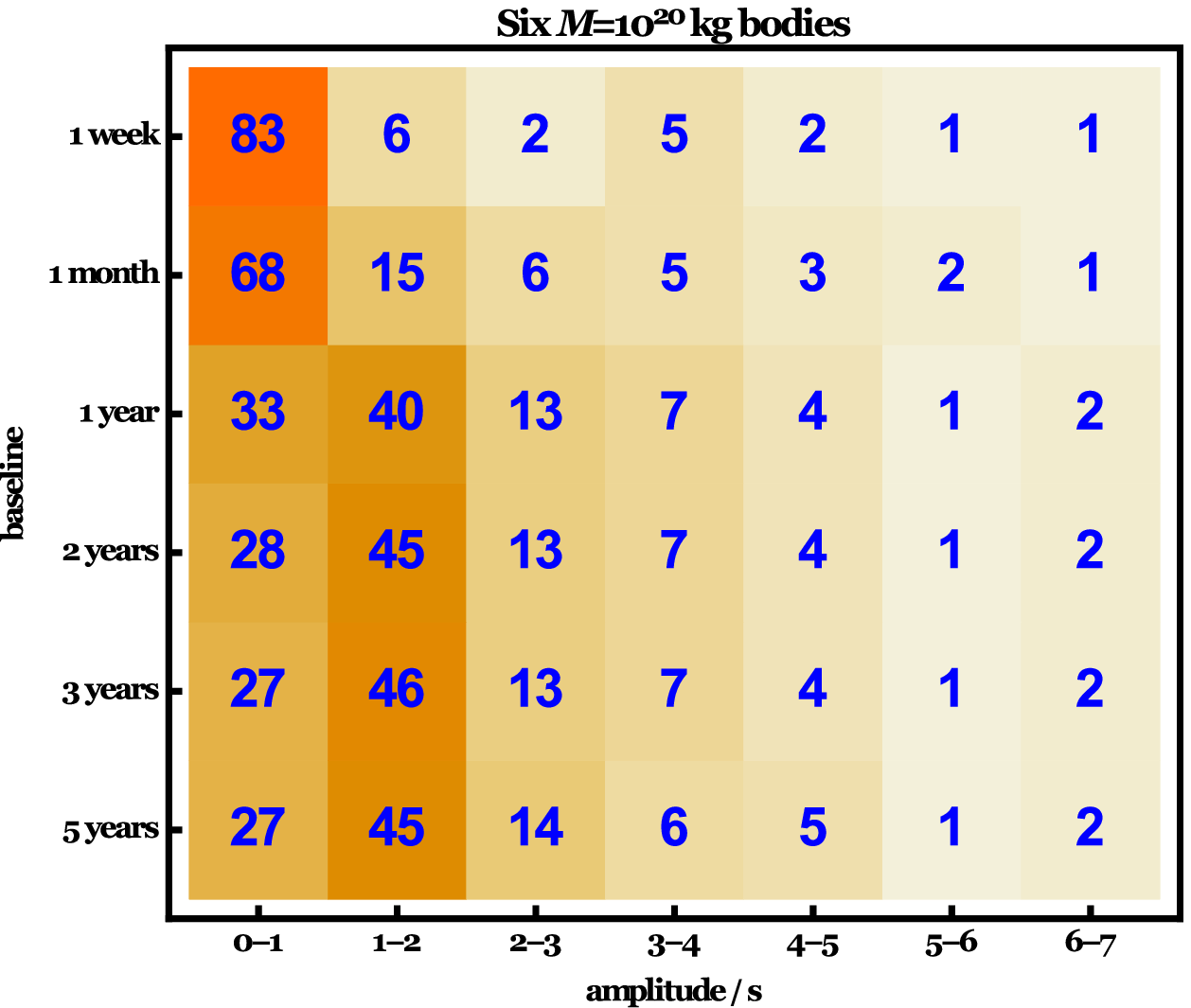}
}
\vspace{0.40em}
\centerline{
\includegraphics[width=7.5cm,height=6.8cm]{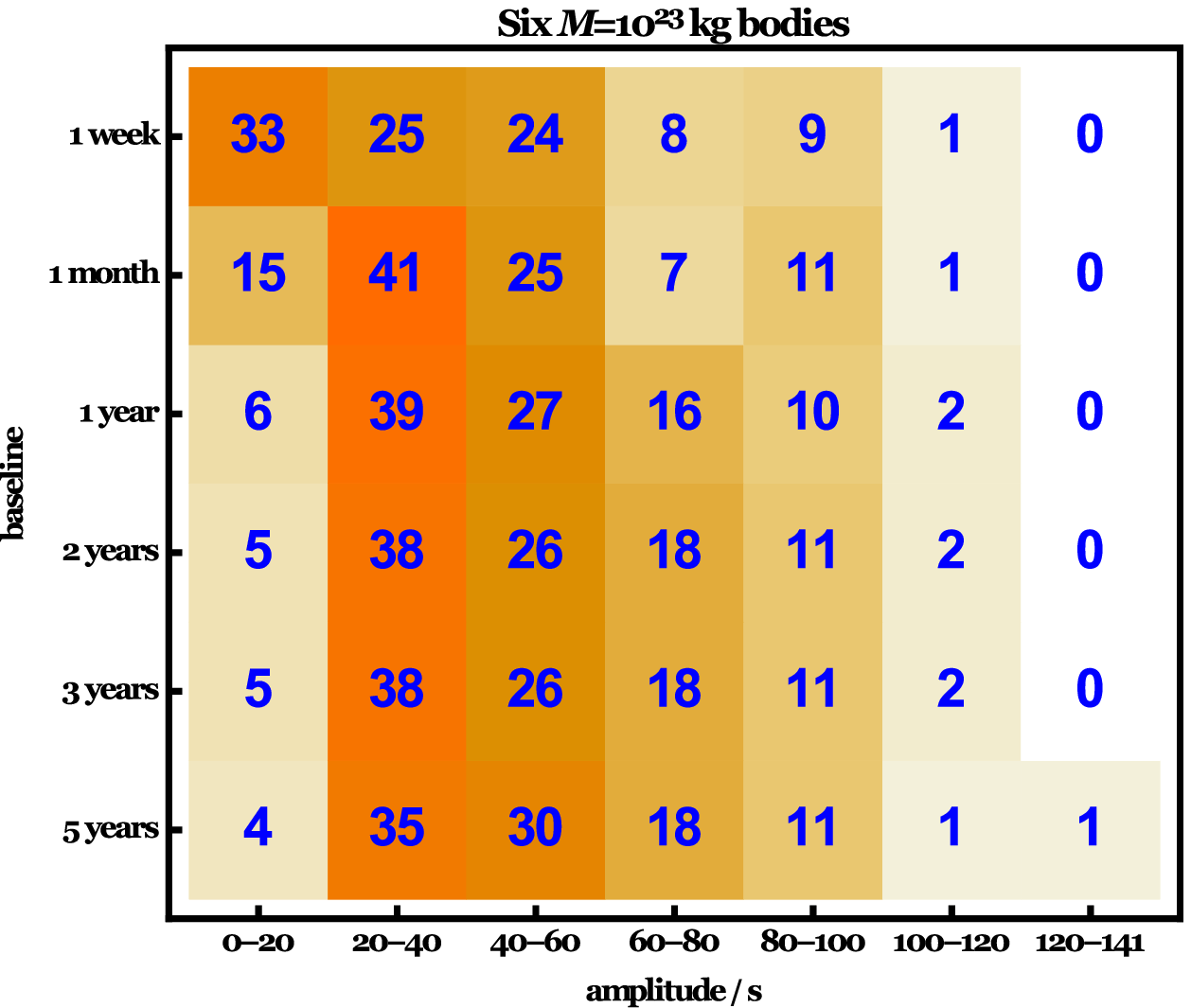}
}
\caption{
The number of systems (embedded blue numbers) for which
a particular maximum orbital period variation
occurred ($x$-axes) over a given baseline of observations
($y$-axes) for suites of 100 stable 
simulations of six co-orbital bodies
with masses $10^{17}$ kg (upper panel),
$10^{20}$ kg (middle panel), and $10^{23}$ kg
(lower panel).  The initial orbital periods of
all bodies are 4.49300 hours.  Both systems from Fig. \ref{linepl}, 
sampled after 5 years, fall in the corresponding 
1-2 second amplitude tile in the middle panel here.
}
\label{hist}
\end{figure}

\begin{figure*}
\centerline{
\includegraphics[width=8cm,height=6.8cm]{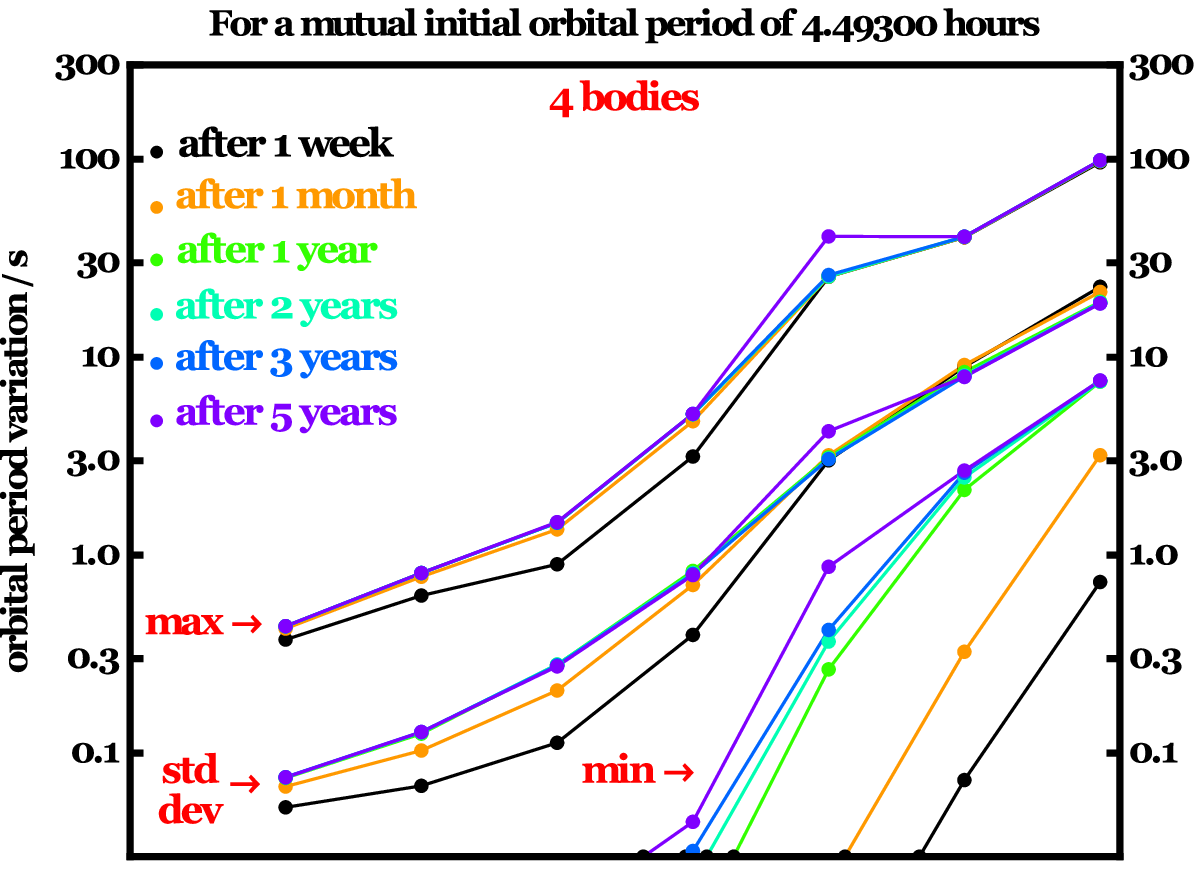}
\includegraphics[width=8cm,height=6.8cm]{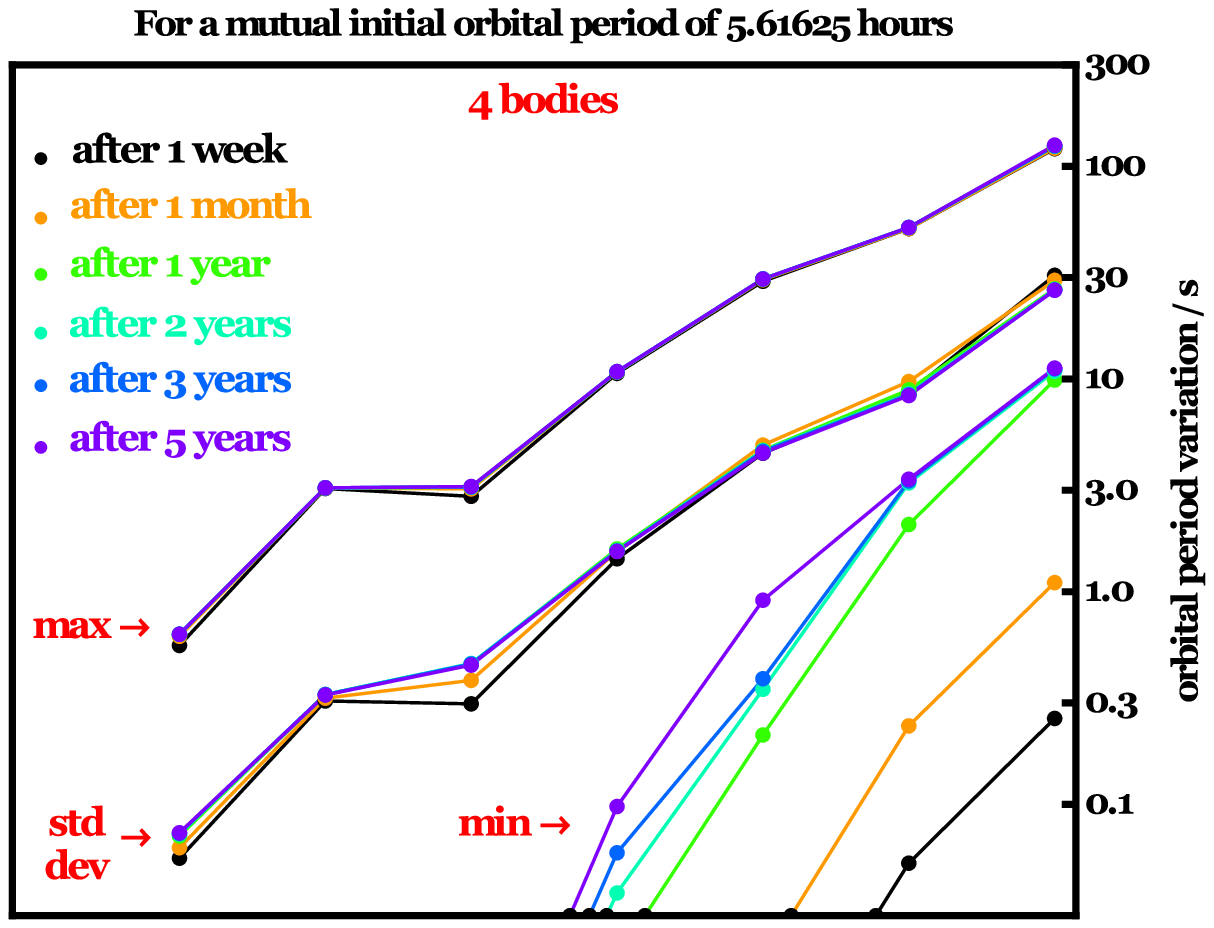}
}
\centerline{
\includegraphics[width=8cm,height=6.8cm]{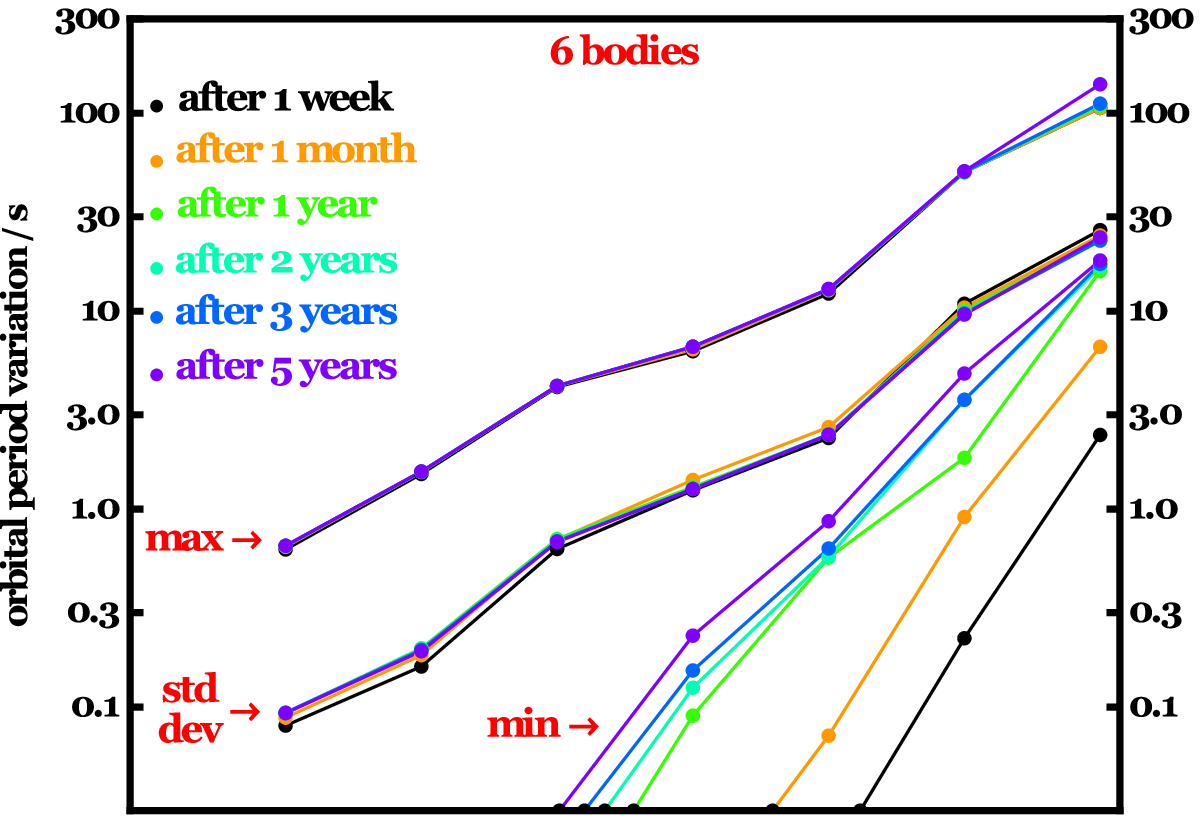}
\includegraphics[width=8cm,height=6.8cm]{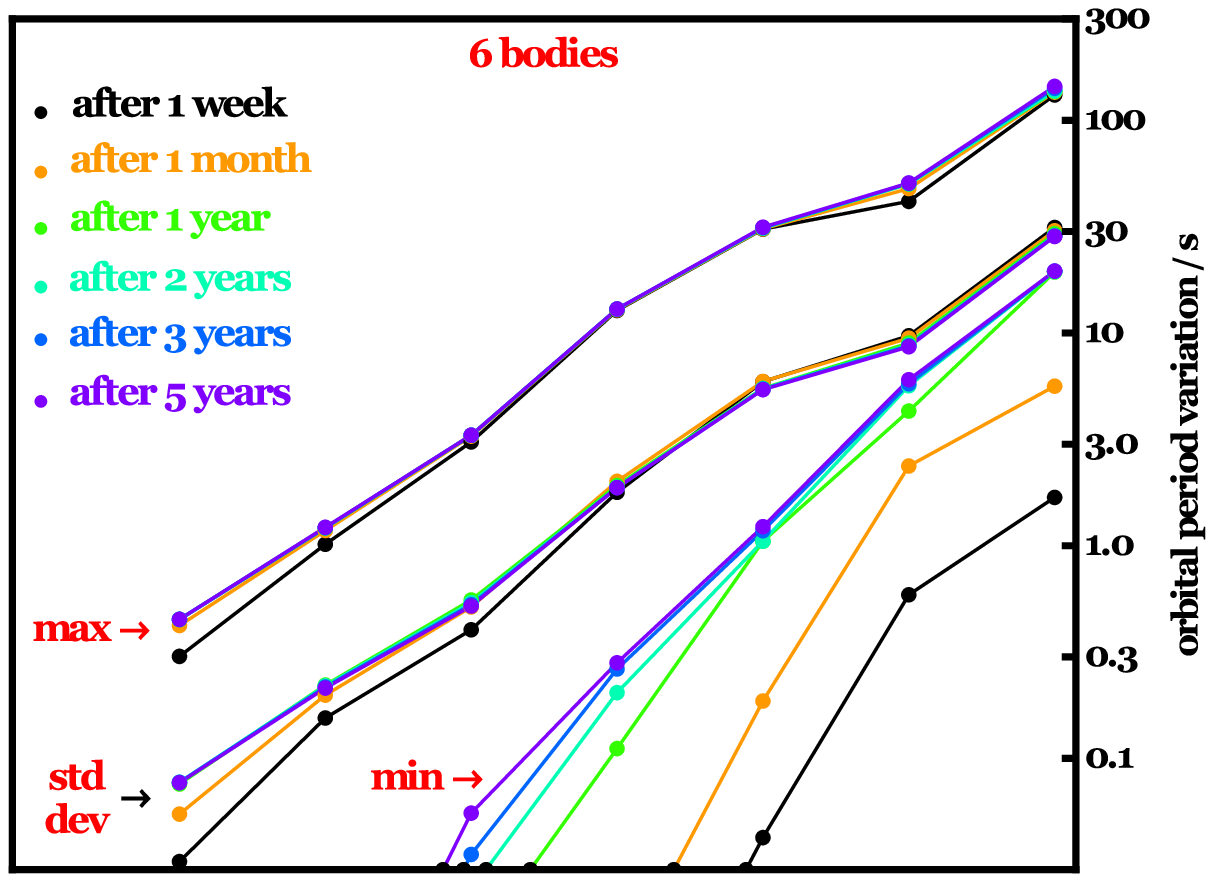}
}
\centerline{
\includegraphics[width=8cm,height=6.8cm]{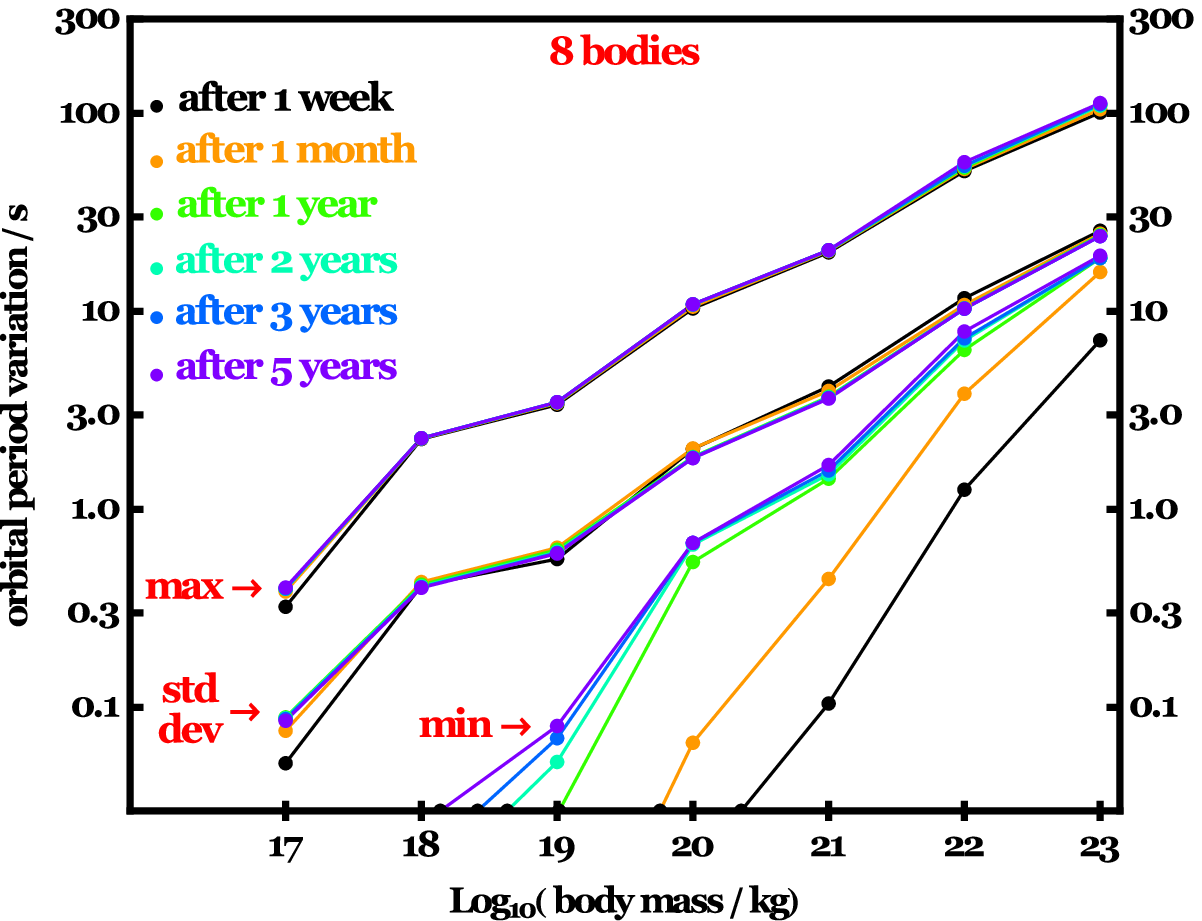}
\includegraphics[width=8cm,height=6.8cm]{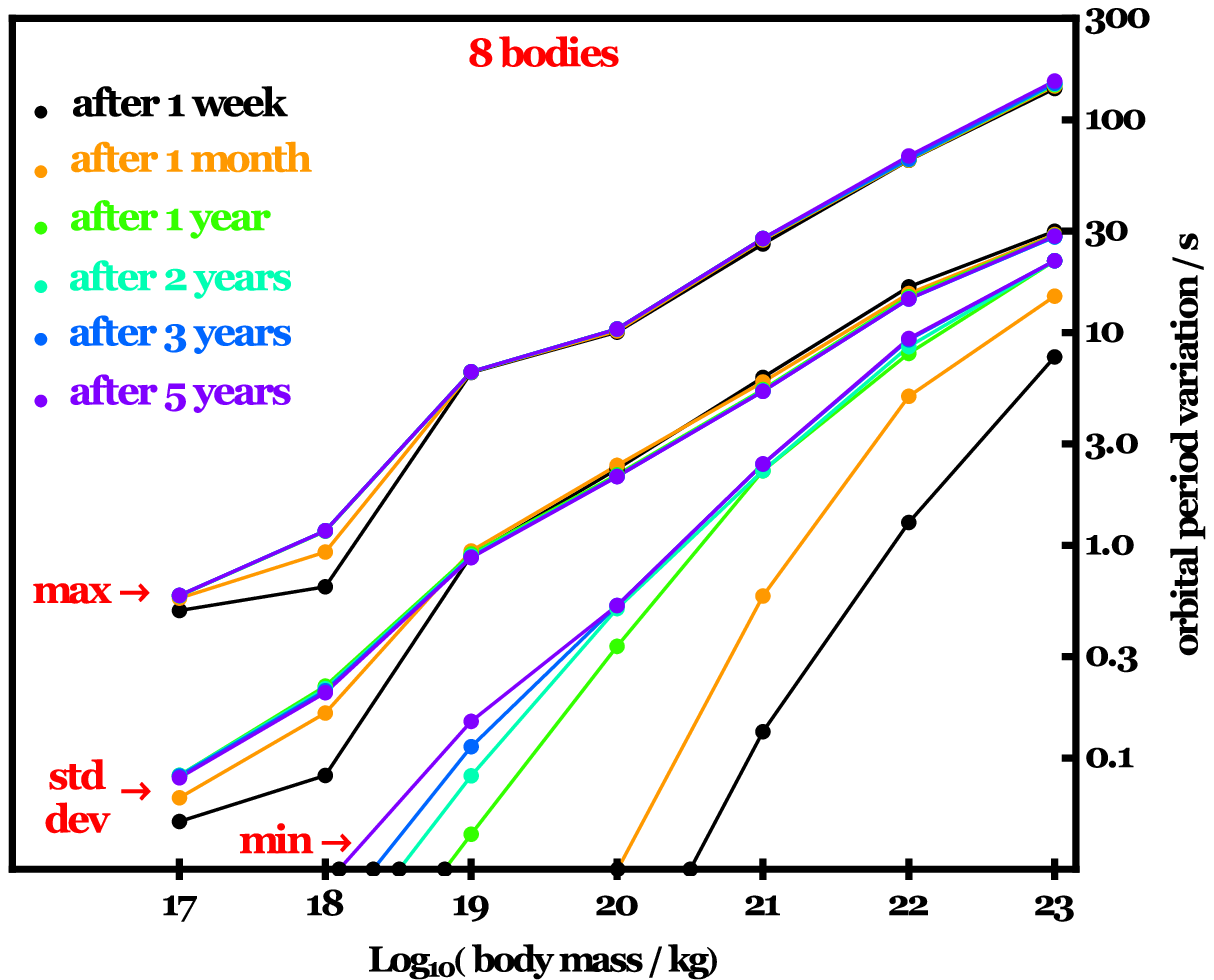}
}
\caption{
Bounding the mass of co-orbital bodies.  
The maximum and minimum orbital period deviations (upper and lower curves, 
respectively, in each plot) for our suites of simulations effectively
provide the lower and upper mass bounds, for a given number of bodies
(4, top; 6, middle; 8, bottom), initial orbital periods (left, 4.49300 hours;
right, 5.61625 hours), and sampling times (labelled different colours).
We also provide the standard deviation of orbital period deviation in
the middle set of curves.  To first order, the plots are insensitive to
multiplicity, and are strongly dependent on co-orbital body mass.
}
\label{Key}
\end{figure*}

\section{Simulation setup}

For our explorations, we use a slightly-modified version of the $N$-body code {\sc Mercury} 
\citep{chambers1999}. The modifications include the effect of general relativity and
better collision detection in the subroutine {\tt mce\_cent}, as was used
in \cite{veretal2013} and \cite{vermus2013}.  These modifications are likely overkill;
general relativity would advance the pericentre of circular co-orbital equal-mass bodies 
equally.  Further, during each orbit, general relativity would cause the bodies to incur 
a maximum inward (non-cumulative) drift of just about 4-6 km \citep{veras2014b}.

We establish our initial conditions according to a scenario where $K$ co-orbital
bodies each of mass $M$ orbit a star of mass $M_{\star}$. Without loss of generality,
we henceforth refer to the star as a white dwarf.  The initial
mutual orbit of the bodies is circular with semimajor axis $a$.  
For each set of $(K,M,M_{\star},a)$, we run over 100 simulations such that each one
features bodies whose initial mean anomalies are drawn from a uniform random
distribution.  We run just enough simulations per set such that exactly 100 remain stable 
for the duration of five years (we also keep and count the unstable simulations for later analysis).  
We output data every 0.25 day, and report orbital period
variations after one week (defined as seven days), one month (defined as 30 days), one year
(defined as 365 days), two years, three years and five years.  

We sample all permutations of $(K,M,M_{\star},a)$ where $K = \left\lbrace 4,6,8 \right\rbrace$ and
$M = \left\lbrace 10^{17}, 10^{18}, 10^{19}, 10^{20}, 10^{21}, 10^{22}, 10^{23} \right\rbrace$ kg.
For the sake of adopting a stellar mass and  period of the co-orbital body,
we adopt the values for WD 1145+017, i.e. $M_{\star}=0.60M_{\odot}$, and $P = 4.49300$ hours,
as well as a value 25\% larger ($P = 5.61625$ hours). We note that the white dwarf
mass in WD1145+017 is not yet accurately known, and that these parameters
are equally illustrative for close-in planets at main sequence stars (e.g.
for KOI 1843.03, with $M_{\star}=0.46M_{\odot}$ and $P = 4.245$ hours; \citealt*{rapetal2013}).
For a given $P$, the values of $a$ change strictly according to what $M$ is being sampled.  

The value of $10^{23}$ kg represents a realistic upper bound for individual masses of co-orbital
bodies which we might expect.  Planet-mass objects are rarely 
thought to enter the white dwarf Roche radius \citep{veretal2013,musetal2014}, particularly
for large planets \citep{vergae2015,veretal2016a}.  The probability, however, increases for 
asteroid-sized \citep{bonetal2011,bonwya2012,debetal2012,frehan2014,bonver2015} or 
moon-sized \citep{payetal2016a,payetal2016b} bodies, given the presence of eccentric planets
\citep{antver2016}; comets enter the Roche radius approximately
once every $10^4$ years \citep{alcetal1986,veretal2014d} but are subject to quick evaporation
\citep{stoetal2015,veretal2015b,broetal2016}.  Further, planets are complex multi-layered objects whose
tidal disruption around main sequence stars \citep{guietal2011,liuetal2013} implies that dedicated treatments
would be necessary for white-dwarf-based studies.

Although all bodies in our simulations are treated with point mass dynamics, we 
give the white dwarf a finite
fiducial radius of 8750 km in order to detect any potential collisions with bodies.
As $M$ increases, we would expect instability to occur on a more frequent basis, and this
instability can manifest as collisions between bodies, collisions between a body and the white dwarf, and ejections.  We set the ejection radius at $3 \times 10^5$ au, which represents
a reasonable upper bound on each axis of the true Hill ellipsoid of planetary systems in the Solar neighbourhood 
\citep{vereva2013,veretal2014c}.  The duration of our simulations (5 yr) ensures that no scattered
object would actually have time to reach this edge, but this value allows any object on its way out to be 
retained and tracked.

\section{Results}

Our simulations yield osculating values of each body's orbital elements, as well as indications
about which have remained stable.  Of greatest interest is the osculating semimajor axis, which can
be converted to orbital period, a direct observable in transiting systems.  Subsections 4.1
and 4.2 showcase the results for orbital period variations and instability, respectively.
Subsection 4.3 then assesses the applicability of these results to the WD 1145+017 system.

\subsection{Orbital period variations} 

Each body varies its orbital period in a non-trivial manner due to the number and distribution
of objects in each system.  Consequently, we rely on statistics to make gross characterisations.  For
an individual observed system, if the phases of each object are known, then a more focused study may be carried 
out.  Such investigations may also consider bodies hidden from view which may significantly contribute
to orbital period variations.

We illustrate two examples of the nonuniformity of the orbital period variations in
Fig. \ref{linepl}.  Both systems in the figure adopt
$(K = 6, M = 10^{20} \ {\rm kg} , M_{\star} = 0.6M_{\odot}, a = 0.00535 \ {\rm au})$
-- where we have used the WD\,1145+017 system as a guide \citep{gaeetal2016,rapetal2016} --
but with different initial mean anomalies (both randomly chosen sets of initial phases).
The top panels show how the amplitude and other properties of these periods vary with time, 
so that the period variation is a function of observing baseline.  The top panels
also illustrate different behaviour from each other: the evolution in the right panel
exhibits quicker amplitude changes from the left panel.

The bottom two panels on each side of the figure are ``O-C'' (Observed - Calculated) diagrams.
They illustrate how, for these two sets of parameters
in particular, the times
of the observed transit dips deviate from a linear dependence on orbital phase (ephemeris).
Observations are more sensitive to relative phase shifts than the actual period
variation, provided that individual transits are unambiguously identified
over a sufficiently long baseline of observations. The middle and bottom panels, respectively, 
illustrate the deviations from linearity in transit time after six months and five years.
The five-year cases represent the full baseline of simulations, whereas the
six-month cases represent perhaps more realistic scenarios.
In all four cases, the deviations are large enough to be detected. The extent
of the deviation can vary by a factor of three even though the initial phases
are similarly clustered (see figure caption).

The matrix plots in Fig. \ref{hist} show how commonly systems with a given
mass of co-orbiting bodies and time baseline achieve particular 
maximum period deviations. The colours indicate the magnitude of the number
of systems out of 100 that achieve amplitudes in a range whose maximum
extent was found to be 0.7 seconds (upper panel,
for $M = 10^{17}$ kg), 7 seconds (middle panel, for $M = 10^{20}$ kg)
and 141 seconds (lower panel, for $M = 10^{23}$ kg).  There does not 
exist one necessarily representative distribution (or
colour scheme) for all sets 
of $(K, M, M_{\star}, a)$.  Outliers are the result of special configurations of
multi-body co-orbital problems, such as $60^{\circ}$ offsets for the much simpler $K=2$
case under the guise of the restricted three-body problem.

In aggregate, however, over the entire phase space there exist clear trends, which are displayed in 
Fig. \ref{Key}. That figure is particularly important because
it effectively bounds the masses from above and below with the bottom and
top sets of curves.  These sets correspond to the minimum and maximum period deviations
obtained in any of the stable simulations that were run; the middle curves display the standard
deviation.  

From these curves, also note 
(i) the clear upward trend in orbital period variation as a function of $M$,
(ii) the slight increase in maximum deviation values for the larger $P$ value
(right panels), and
(iii) the upper and middle sets of curves in each plot are 
largely insensitive to the sampling timescale, unlike
the lower curves. This trend may be inferred from Fig.\,\ref{linepl}.

\begin{figure}
\centerline{
\includegraphics[width=8cm,height=6cm]{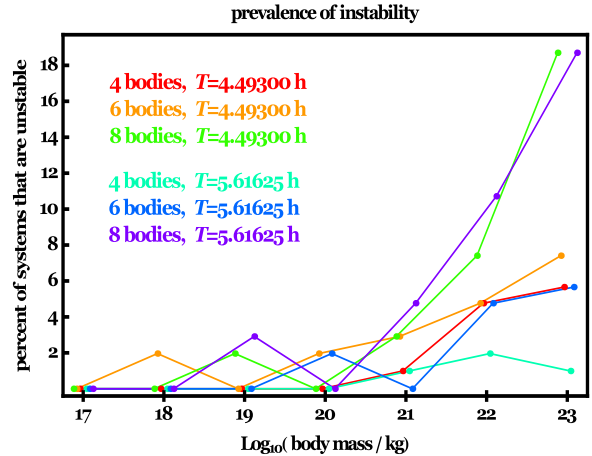}
}
\caption{
Fraction of unstable systems for different combinations of 
multiplicity, mass and orbital period.  Each data point is slightly 
horizontally offset for clarity.  In at least
95\%, 90\% and 80\% of all simulations, masses under 
$10^{21}, 10^{22}$ and $10^{23}$ kg respectively kept the systems stable.
Consequently, in general terms, instability does not provide as robust a
bound on the masses as does the maximum and minimum period deviations
(Fig. \ref{Key}).
}
\label{inst}
\end{figure}

\subsection{Instability}  

One potential constraint on the upper mass bound of bodies is system
stability. Although instability may come in the form of engulfments within the star,
bodies being perturbed on a course out of the system, or collisions amongst bodies, only 
the last possibility
occurred in our simulations.  The likely reason is because any close encounters 
between the bodies, given their masses, could not have generated the speed needed
to eventually escape the system. Further, co-orbital circular configurations are geometrically unfavoured
to produce ejections or engulfments within the parent star. Therefore, even if we had instead adopted
a stellar radius that was commensurate with a small main sequence star, engulfments would
still not have been likely to occur unless the co-orbital bodies were large enough to be affected by star-planet tides.

There does not exist a particular critical body mass at which a system would go unstable
because stability is a function of both $K$ and $M$ for a fixed number of orbits.  Nevertheless, 
the probability of instability increases as both $K$ and $M$ do, as demonstrated by Fig. \ref{inst}.
The plot makes clear predictions that instability should occur under the 5\%, 10\% and 20\%
level for $M \le 10^{21}, 10^{22}$ and $10^{23}$ kg.  Regarding bounding body masses from observations,
these constraints are weak compared to those obtained from orbital period deviations.

Finally, we note that collisions amongst bodies might be observable.  The
result will be a reduced value of $K$, and a change in orbital period variations.
In this scenario, however, the masses of the remaining co-orbital bodies are no
longer likely to be equal, and consequently a more detailed analysis would 
be required.

\subsection{Comparison to WD 1145+017}

As suggested in the introduction, the WD 1145+017 system is too complex to
be modelled by the architecture and masses that we have adopted here.
Nevertheless, one interesting comparison can be made: typical orbital period
deviations that are reported in this study for $M = 10^{20}$ kg are roughly a few seconds
(Fig. \ref{Key}). \cite{rapetal2016} deduced that an asteroid mass of this order of magnitude
can, through fragmentation, produce bodies which settle into an orbit whose period
deviates from the original orbit by about 27 seconds. This larger, observed, deviation in WD 1145+017
may easily be explained by effects not accounted for here, which include
(i) the orbiting objects are probably of different mass, (ii) the process of fragmentation,
(iii) the known presence of dust, and (iv) the known presence of gas.

\section{Summary}

Multiple co-orbital bodies near or
inside of a stellar Roche radius might provide unique insights
into a system's violent history. For white dwarfs, these
bodies provide a glimpse into how debris discs are formed and 
how white dwarf atmospheres are polluted.  The mutual perturbations amongst the bodies
generate slight variations in orbital periods ($\sim 0.1-100$ s)
and phase shifts over the course of weeks, months or years, which are detectable by 
current instruments.  Here, we have quantified this variation as
a function of mass, multiplicity, distance and time sampling 
(Fig. \ref{Key}) and show that for a given orbital period deviation,
lower and upper mass bounds may be estimated.  We also characterised the fuzzy
instability boundary (Fig. \ref{inst}), which provides an upper bound
on the incidence of instability for a given co-orbital body mass.

\section*{Acknowledgements}

We thank the referee for their useful feedback on this manuscript.
DV and BTG have received funding from the European Research Council under the European Union's Seventh Framework
Programme (FP/2007-2013)/ERC Grant Agreement n. 320964 (WDTracer).  TRM was supported under a Science and 
Technology Facilities Council (STFC) grant, ST/L000733.

\label{lastpage}

\begin{thebibliography}{99}

\bibitem[Alcock et al.(1986)]{alcetal1986} Alcock, C., Fristrom, C.~C., \& Siegelman, R.\ 1986, ApJ, 302, 462 

\bibitem[Alonso et al.(2016)]{aloetal2016} Alonso, R., Rappaport, S., Deeg, H.~J., \& Palle, E.\ 2016, In Press, A\&A, arXiv:1603.08823 

\bibitem[Antoniadou et al.(2014)]{antetal2014} Antoniadou, K.~I., Voyatzis, G., \& Varvoglis, H.\ 2014, IAU Symposium, 310, 82 

\bibitem[Antoniadou \& Veras(2016)]{antver2016} Antoniadou, K.~I., Veras, D.\ 2016, Submitted to MNRAS 

\bibitem[Barber et al.(2016)]{baretal2016} Barber, S.~D., Belardi, C., Kilic, M., \& Gianninas, A.\ 2016, MNRAS, 459, 1415 

\bibitem[Bear \& Soker(2013)]{beasok2013} Bear, E., \& Soker, N.\ 2013, New Astronomy, 19, 56

\bibitem[Bear \& Soker(2015)]{beasok2015} Bear, E., \& Soker, N.\ 2015, MNRAS, 450, 4233 

\bibitem[Beaug{\'e} et al.(2007)]{beaetal2007} Beaug{\'e}, C., S{\'a}ndor, Z., {\'E}rdi, B., S\"uli, {\'A}.\ 2007, A\&A, 463, 359 

\bibitem[Bengochea et al.(2015)]{benetal2015} Bengochea, A., 
Gal{\'a}n, J., \& P{\'e}rez-Chavela, E.\ 2015, Physica D Nonlinear Phenomena, 301, 21 

\bibitem[Bergfors et al.(2014)]{beretal2014} Bergfors, C., Farihi, 
J., Dufour, P., \& Rocchetto, M.\ 2014, MNRAS, 444, 2147

\bibitem[Bochinski et al.(2015)]{bocetal2015} Bochinski, J.~J., Haswell, C.~A., Marsh, T.~R., Dhillon, V.~S., \& Littlefair, S.~P.\ 2015, ApJL, 800, L21 

\bibitem[Bonsor et al.(2011)]{bonetal2011} Bonsor, A., Mustill, A.~J., \& Wyatt, M.~C.\ 2011, MNRAS, 414, 930 

\bibitem[Bonsor \& Wyatt(2012)]{bonwya2012} Bonsor, A., \& Wyatt, M.~C.\ 2012, MNRAS, 420, 2990 

\bibitem[Bonsor \& Veras(2015)]{bonver2015} Bonsor, A., \& Veras, D.\ 2015, MNRAS, 454, 53 

\bibitem[Brown, Veras \& G{\"a}nsicke(2016)]{broetal2016} Brown, J.~C., Veras, D., G{\"a}nsicke, B.~T.\ 2016, In prep 

\bibitem[Chambers(1999)]{chambers1999} Chambers, J.~E.\ 1999, MNRAS, 304, 793 

\bibitem[Cordes \& Shannon(2008)]{corsha2008} Cordes, J.~M., \& Shannon, R.~M.\ 2008, ApJ, 682, 1152 

\bibitem[Cresswell \& Nelson(2009)]{crenel2009} Cresswell, P., \& Nelson, R.~P.\ 2009, A\&A, 493, 1141 

\bibitem[Croll et al.(2014)]{croetal2014} Croll, B., Rappaport, S., DeVore, J., et al.\ 2014, ApJ, 786, 100 

\bibitem[Croll et al.(2015)]{croetal2015} Croll, B., Dalba, P.~A., 
Vanderburg, A., et al.\ 2015, arXiv:1510.06434 

\bibitem[Debes et al.(2012)]{debetal2012} Debes, J.~H., Walsh, K.~J., \& Stark, C.\ 2012, ApJ, 747, 148

\bibitem[Dobrovolskis(2013)]{dobrovolskis2013} Dobrovolskis, A.~R.\ 2013, Icarus, 226, 1635 

\bibitem[Dollfus(1967)]{dollfus1967} Dollfus, A.\ 1967, Sky \& Telescope, 34,  

\bibitem[Farihi et al.(2009)]{faretal2009} Farihi, J., Jura, M., \& Zuckerman, B.\ 2009, ApJ, 694, 805 

\bibitem[Farihi(2016)]{farihi2016} Farihi, J. Submitted to New Astronomy Reviews

\bibitem[Ford \& Gaudi(2006)]{forgau2006} Ford, E.~B., \& Gaudi, B.~S.\ 2006, ApJL, 652, L137 

\bibitem[Ford \& Holman(2007)]{forhol2007} Ford, E.~B., \& Holman, M.~J.\ 2007, ApJL, 664, L51 

\bibitem[Fountain \& Larson(1978)]{foular1978} Fountain, J.~W., \& Larson, S.~M.\ 1978, Icarus, 36, 92 

\bibitem[Frewen \& Hansen(2014)]{frehan2014} Frewen, S.~F.~N., \& Hansen, B.~M.~S.\ 2014, MNRAS, 439, 2442 

\bibitem[G{\"a}nsicke et al.(2006)]{gaeetal2006} G{\"a}nsicke, 
B.~T., Marsh, T.~R., Southworth, J., 
\& Rebassa-Mansergas, A.\ 2006, Science, 314, 1908 

\bibitem[G{\"a}nsicke et al.(2012)]{ganetal2012} G{\"a}nsicke, 
B.~T., Koester, D., Farihi, J., et al.\ 2012, MNRAS, 424, 333 

\bibitem[G{\"a}nsicke et al.(2016)]{gaeetal2016} G{\"a}nsicke, 
B.~T., Aungwerojwit, A., Marsh, T.~R., et al.\ 2016, ApJL, 818, L7 

\bibitem[Gascheau(1843)]{gascheau1843} Gascheau, G.\ 1843, C.R. Acad. Sci. Paris, 16, 393

\bibitem[Giuppone et al.(2012)]{giuetal2012} Giuppone, C.~A., Ben{\'{\i}}tez-Llambay, P., \& Beaug{\'e}, C.\ 2012, MNRAS, 421, 356 

\bibitem[Go{\'z}dziewski \& Konacki(2006)]{gozkon2006} Go{\'z}dziewski, K., \& Konacki, M.\ 2006, ApJ, 647, 573 

\bibitem[Graham et al.(1990)]{graetal1990} Graham, J.~R., Matthews, K., Neugebauer, G., \& Soifer, B.~T.\ 1990, ApJ, 357, 216 

\bibitem[Guillochon et al.(2011)]{guietal2011} Guillochon, J., Ramirez-Ruiz, E., \& Lin, D.\ 2011, ApJ, 732, 74 

\bibitem[Hadjidemetriou et al.(2009)]{hadetal2009} Hadjidemetriou, J.~D., Psychoyos, D., \& Voyatzis, G.\ 2009, Celestial Mechanics and Dynamical Astronomy, 104, 23 

\bibitem[Hadjidemetriou \& Voyatzis(2011)]{hadvoy2011} Hadjidemetriou, J.~D., \& Voyatzis, G.\ 2011, Celestial Mechanics and Dynamical Astronomy, 111, 179 

\bibitem[Izidoro et al.(2010)]{izoetal2010} Izidoro, A., Winter, O.~C., \& Tsuchida, M.\ 2010, MNRAS, 405, 2132 

\bibitem[Janson(2013)]{janson2013} Janson, M.\ 2013, ApJ, 774, 156

\bibitem[Jura(2003)]{jura2003} Jura, M.\ 2003, ApJL, 584, L91 

\bibitem[Jura \& Young(2014)]{juryou2014} Jura, M., 
\& Young, E.~D.\ 2014, Annual Review of Earth and Planetary Sciences, 42, 45

\bibitem[Klein et al.(2011)]{kleetal2011} Klein, B., Jura, M., 
Koester, D., \& Zuckerman, B.\ 2011, ApJ, 741, 64 

\bibitem[Koester et al.(2014)]{koeetal2014} Koester, D., G{\"a}nsicke, B.~T., \& Farihi, J.\ 2014, A\&A, 566, A34 

\bibitem[Kortenkamp et al.(2004)]{koretal2004} Kortenkamp, S.~J., Malhotra, R., \& Michtchenko, T.\ 2004, Icarus, 167, 347 

\bibitem[Lagrange(1772)]{lagrange1772} Lagrange, J-L.\ 1772 \OE uvres compl\`etes (Paris: Gouthier-Villars)

\bibitem[Laughlin \& Chambers(2002)]{laucha2002} Laughlin, G., \& Chambers, J.~E.\ 2002, AJ, 124, 592 

\bibitem[Leleu et al.(2015)]{leletal2015} Leleu, A., Robutel, P., \& Correia, A.~C.~M.\ 2015, A\&A, 581, A128 

\bibitem[Liu et al.(2013)]{liuetal2013} Liu, S.-F., Guillochon, J., Lin, D.~N.~C., \& Ramirez-Ruiz, E.\ 2013, ApJ, 762, 37 

\bibitem[Manser et al.(2016)]{manetal2016} Manser, C.~J.
G{\"a}nsicke, B.~T., Marsh, T.~R., et al.\ 2016, MNRAS, 455, 4467 

\bibitem[Maxwell(1890)]{maxwell1890} Maxwell, C.~J.\ 1890 On the stability of the motion of Saturn's Rings, in Scientific Papers of James Clerk Maxwell, Cambridge University Press, Vol 1, 228.

\bibitem[Mayor \& Queloz(1995)]{mayque1995} Mayor, M., \& Queloz, D.\ 1995, Nature, 378, 355 

\bibitem[Metzger et al.(2012)]{metetal2012} Metzger, B.~D., Rafikov, R.~R., \& Bochkarev, K.~V.\ 2012, MNRAS, 423, 505 

\bibitem[Moeckel(1994)]{moeckel1994} Moeckel, R.\ 1994, J. Dyn. Diff. Eq., 6(1), 35

\bibitem[Murray \& Dermott(1999)]{murder1999} Murray, C.~D., \& Dermott, S.~F.\ 1999, Solar system dynamics  

\bibitem[Mustill et al.(2014)]{musetal2014} Mustill, A.~J., Veras, D., \& Villaver, E.\ 2014, MNRAS, 437, 1404 

\bibitem[Nauenberg(2002)]{nauenberg2002} Nauenberg, M.\ 2002, AJ, 124, 2332

\bibitem[Payne et al.(2016a)]{payetal2016a} Payne, M.~J., Veras, D., Holman, M.~J., G\"{a}nsicke, B.~T.\ 2016a, MNRAS, 457, 217 

\bibitem[Payne et al.(2016b)]{payetal2016b} Payne, M.~J., Veras, D., G\"{a}nsicke, B.~T., Holman, M.~J., \ 2016b, Submitted to MNRAS 

\bibitem[Pendse(1935)]{pendse1935} Pendse, C.~G.\ 1935, Phil. Trans. Roy. Soc. London (A), 234, 145

\bibitem[Pierens \& Raymond(2014)]{pieray2014} Pierens, A., \& Raymond, S.~N.\ 2014, MNRAS, 442, 2296 

\bibitem[Placek et al.(2015)]{plaetal2015} Placek, B., Knuth, K.~H., Angerhausen, D., \& Jenkins, J.~M.\ 2015, ApJ, 814, 147 

\bibitem[Rafikov(2011a)]{rafikov2011a} Rafikov, R.~R.\ 2011a, MNRAS, 416, L55

\bibitem[Rafikov(2011b)]{rafikov2011b} Rafikov, R.~R.\ 2011b, ApJL, 732, LL3

\bibitem[Rafikov \& Garmilla(2012)]{rafgar2012} Rafikov, R.~R., \& Garmilla, J.~A.\ 2012, ApJ, 760, 123

\bibitem[Rappaport et al.(2012)]{rapetal2012} Rappaport, S., Levine, A., Chiang, E., et al.\ 2012, ApJ, 752, 1 

\bibitem[Rappaport et al.(2013)]{rapetal2013} Rappaport, S., Sanchis-Ojeda, R., Rogers, L.~A., Levine, A., \& Winn, J.~N.\ 2013, ApJ, 773, L15 

\bibitem[Rappaport et al.(2014)]{rapetal2014} Rappaport, S., Barclay, T., DeVore, J., et al.\ 2014, ApJ, 784, 40 

\bibitem[Rappaport et al.(2016)]{rapetal2016} Rappaport, S., Gary, 
B.~L., Kaye, T., et al.\ 2016, arXiv:1602.00740 

\bibitem[Renner \& Sicardy(2004)]{rensic2004} Renner, S., \& Sicardy, B.\ 2004, 
Celestial Mechanics and Dynamical Astronomy, 88, 397 

\bibitem[Robutel \& Pousse(2013)]{robpou2013} Robutel, P., \& Pousse, A.\ 2013, Celestial Mechanics and Dynamical Astronomy, 117, 17 

\bibitem[Routh(1875)]{routh1875} Routh, E.J.\ 1875, Proc. Lond. Math. Soc., 6, 86 

\bibitem[Salo \& Yoder(1988)]{salyod1988} Salo, H., \& Yoder, C.~F.\ 1988, A\&A, 205, 309 

\bibitem[Sanchis-Ojeda et al.(2015)]{sanetal2015} Sanchis-Ojeda, R., Rappaport, S., Pall{\`e}, E., et al.\ 2015, ApJ, 812, 112 

\bibitem[Schwarz et al.(2005)]{schetal2005} Schwarz, R., Pilat-Lohinger, E., Dvorak, R., {\'E}rdi, B., 
\& S{\'a}ndor, Z.\ 2005, Astrobiology, 5, 579 

\bibitem[Schwarz et al.(2015)]{schetal2015} Schwarz, R., Bazs{\'o}, {\'A}., Funk, B., \& Zechner, R.\ 2015, 
MNRAS, 453, 2308 

\bibitem[Smith \& Lissauer(2010)]{smilis2010} Smith, A.~W., \& Lissauer, J.~J.\ 2010, Celestial Mechanics and Dynamical Astronomy, 107, 487 

\bibitem[Stone et al.(2015)]{stoetal2015} Stone, N., Metzger, B.~D., \& Loeb, A.\ 2015, MNRAS, 448, 188 

\bibitem[Tremblay et al.(2016)]{treetal2016} Tremblay, P.-E., et al., Submitted to MNRAS

\bibitem[Vanderburg et al.(2015)]{vanetal2015} Vanderburg, A., 
Johnson, J.~A., Rappaport, S., et al.\ 2015, Nature, 526, 546 

\bibitem[Veras et al.(2013)]{veretal2013} Veras, D., Mustill, A.~J., Bonsor, A., \& Wyatt, M.~C.\ 2013, MNRAS, 431, 1686 

\bibitem[Veras \& Evans(2013)]{vereva2013} Veras, D., \& Evans, N.~W.\ 2013, MNRAS, 430, 403 

\bibitem[Veras \& Mustill(2013)]{vermus2013} Veras, D., \& Mustill, A.~J.\ 2013, MNRAS, 434, L11 

\bibitem[Veras(2014a)]{veras2014a} Veras, D.\ 2014a, Celestial Mechanics and Dynamical Astronomy, 118, 315 

\bibitem[Veras(2014b)]{veras2014b} Veras, D.\ 2014b, MNRAS, 442, L71 

\bibitem[Veras et al.(2014a)]{veretal2014a} Veras, D., Leinhardt, Z.~M., Bonsor, A., G\"{a}nsicke, B.~T.\ 2014a, MNRAS, 445, 2244 

\bibitem[Veras et al.(2014b)]{veretal2014b} Veras, D., Jacobson, S.~A., G\"{a}nsicke, B.~T.\ 2014b, MNRAS, 445, 2794

\bibitem[Veras et al.(2014c)]{veretal2014c} Veras, D., Evans, N.~W., Wyatt, M.~C., \& Tout, C.~A.\ 2014c, MNRAS, 437, 1127 

\bibitem[Veras et al.(2014d)]{veretal2014d} Veras, D., Shannon, A., G\"{a}nsicke, B.~T.\ 2014d, MNRAS, 445, 4175 

\bibitem[Veras \& G\"{a}nsicke(2015)]{vergae2015} Veras, D., G\"{a}nsicke, B.~T.\ 2015, MNRAS, 447, 1049 

\bibitem[Veras et al.(2015a)]{veretal2015a} Veras, D., Eggl, S., G\"{a}nsicke, B.~T.\ 2015a, MNRAS, 451, 2814 

\bibitem[Veras et al.(2015b)]{veretal2015b} Veras, D., Eggl, S., G\"{a}nsicke, B.~T.\ 2015b, MNRAS, 452, 1945

\bibitem[Veras et al.(2015c)]{veretal2015c} Veras, D., Leinhardt, Z.~M., Eggl, S., G\"{a}nsicke, B.~T.\ 2015c, MNRAS, 451, 3453 

\bibitem[Veras(2016)]{veras2016} Veras, D.\ 2016, Royal Society Open Science. 3:150571

\bibitem[Veras et al.(2016a)]{veretal2016a} Veras, D., Mustill, A.~M., G\"{a}nsicke, B.~T., Redfield, S., Georgakarakos, N., Bowler, A.~B., Lloyd, M.~J.~S.\ 2016a, Submitted to MNRAS

\bibitem[Vokrouhlick{\'y} \& Nesvorn{\'y}(2014)]{voknes2014} Vokrouhlick{\'y}, D., \& Nesvorn{\'y}, D.\ 2014, ApJ, 791, 6 

\bibitem[Wilson et al.(2014)]{wiletal2014} Wilson, D.~J., 
G{\"a}nsicke, B.~T., Koester, D., et al.\ 2014, MNRAS, 445, 1878 

\bibitem[Wilson et al.(2015)]{wiletal2015} Wilson, D.~J., 
G{\"a}nsicke, B.~T., Koester, D., et al.\ 2015, MNRAS, 451, 3237

\bibitem[Wilson et al.(2016)]{wiletal2016} Wilson, D.~J., G{\"a}nsicke, B.~T., 
Farihi, J., \& Koester, D.\ 2016, MNRAS, 459, 3282 

\bibitem[Wolf(1906)]{wolf1906} Wolf, M.\ 1906, Astron. Nachr., 170, 353

\bibitem[Wolszczan \& Frail(1992)]{wolfra1992} Wolszczan, A., \& Frail, D.~A.\ 
1992, Nature, 355, 145 

\bibitem[Wolszczan(1994)]{wolszczan1994} Wolszczan, A.\ 1994, Science, 264, 538 
wolszczan

\bibitem[Xu et al.(2014)]{xuetal2014} Xu, S., Jura, M., Koester, 
D., Klein, B., \& Zuckerman, B.\ 2014, ApJ, 783, 79 

\bibitem[Xu et al.(2016)]{xuetal2016} Xu, S., Jura, M., Dufour, 
P., \& Zuckerman, B.\ 2016, ApJL, 816, L22 

\bibitem[Zhou et al.(2016)]{zhoetal2016} Zhou, G., Kedziora-Chudczer, L., Bailey, J., et al.\ 2016, 
Submitted to MNRAS, arXiv:1604.07405 

\bibitem[Zuckerman \& Becklin(1987)]{zucbec1987} Zuckerman, B., \& Becklin, E.~E.\ 1987, Nature, 330, 138 

\bibitem[Zuckerman et al.(2003)]{zucetal2003} Zuckerman, B., Koester, D., Reid, I.~N., H\"{u}nsch, M.\ 2003, ApJ, 596, 477 

\bibitem[Zuckerman et al.(2007)]{zucetal2007} Zuckerman, B., Koester, D., Melis, C., Hansen, B.~M., \& Jura, M.\ 2007, ApJ, 671, 872 

\bibitem[Zuckerman et al.(2010)]{zucetal2010} Zuckerman, B., Melis, C., Klein, B., Koester, D., \& Jura, M.\ 2010, ApJ, 722, 725 


\end{thebibliography}
\end{document}